\documentstyle[psfig,aps,prl,amsfonts,amssymb]{revtex}
\begin{document}
\title{{\it a priori} Probability  that a Qubit-Qutrit Pair is  Separable}
\author{Paul B. Slater}
\address{ISBER, University of
California, Santa Barbara, CA 93106-2150\\
e-mail: slater@itp.ucsb.edu,	
FAX: (805) 893-7995}

\date{\today}

\draft
\maketitle
\vskip -0.1cm

\begin{abstract}
We extend to arbitrarily coupled pairs of {\it qubits} 
(two-state quantum systems) and {\it qutrits} (three-state quantum systems)  
our earlier study 
({\it Quant. Info. Proc.} [Oct. 2002]), 
which was concerned with the simplest instance of
entangled quantum systems, pairs of qubits. 
As in that analysis --- again on the basis of numerical 
(quasi-Monte Carlo) integration results, but now in a 
still higher-dimensional
space ($35$-d {\it vs}. $15$-d) --- we examine a
conjecture that the Bures/SD (statistical distinguishability) 
probability that arbitrarily paired qubits 
and  qutrits  are
separable (unentangled) has a simple {\it exact} value, $u/(v
 \pi^{3}) 
\approx .00124706$, where $u=2^{20} \cdot 3^3 \cdot 5 \cdot 7$ and
$v=19 \cdot 23 \cdot 29 \cdot 31 \cdot 37 \cdot 41 \cdot 43$
(the product of consecutive primes.) 
 This is considerably less than the conjectured 
value of the Bures/SD probability,
$8 /(11 \pi^2) \approx.0736881$, 
in  the qubit-qubit case. Both of these conjectures, in turn, rely upon ones
to the effect that the SD volumes of separable states assume
certain
remarkable forms, involving ``primorial'' numbers. 
We also estimate the SD {\it area} of the boundary
of separable qubit-qutrit states, and provide preliminary calculations
of the Bures/SD probability of separability in
the general qubit-qubit-qubit and qutrit-qutrit cases.
\end{abstract}

\vspace{.1cm}

\hspace{1cm} {\bf{Key words:}} Qubits, Qutrits, $SU(6)$, 
density matrix, Bures
metric, statistical distinguishability metric, entanglement, Euler 
angle parameterization,  
negativity, concurrence, quasi-Monte Carlo, 
scrambled Halton sequences, numerical integration, primorials

\vspace{.15cm}

In this report, we  
extend to the case of a qubit-{\it qutrit} pair, a form of analysis 
we have previously applied \cite{slaterpreprint} to a pair of quantum bits 
(qubits). 
In that study, on the basis of numerical (quasi-Monte Carlo) integration
results, coupled with earlier {\it exact} findings in related
(lower-dimensional) more specialized 
contexts \cite{slater2} (cf. \cite{slaterJPA}), 
we formulated a 
conjecture that the {\it a priori} (Bures/statistical 
distinguishability [SD]) probability that two qubits are separable is 
\begin{equation} \label{e1}
 P_{4} = {2^3 \over 11 \pi^2}  \approx .0736881.
\end{equation}

The Bures probability that was employed 
was taken to be the normalized {\it volume element}
of the Bures {\it metric}. This metric 
 has been the subject of considerable
study in the past few decades, starting with the work of Bures
himself \cite{bures}. Then, Uhlmann --- analyzing transition
probabilities in *-algebras \cite{uhlmann}  --- and a number of his students
(M. H\"ubner, G. Rudolph, J. Dittmann,...)  at
the University of Leipzig investigated a wide range of 
interesting properties of the metric. For example,
Dittmann gave a certain 
Yang-Mills characterization \cite{ditt1} (cf. \cite{slaterYM}).
 He also developed
pertinent computational procedures \cite{ditt2}, which have been utilized by
Slater in a number of studies \cite{slater2,slaterJPA,slateregg}.
 Braunstein and Caves \cite{bc},
by finding measurements that optimally resolved neighboring quantum states,
used statistical distinguishability to define a natural Riemannian
(that is, the Bures) metric on the space of quantum-mechanical density 
operators. Also,
Fuchs and Caves gave a statistically motivated derivation of the
Bures-Uhlmann measure of distinguishability for density operators \cite{fuchs}. Petz and Sud\'ar extensively discussed the class of {\it monotone}
metrics, in which the Bures metric plays the {\it minimal} 
role \cite{petz}. \.Zyczkowski, Horodecki, Sanpera, and Lewenstein
\cite{zhsl} gave ``philosophical,'', ``practical'' and ``physical''
reasons for studying the relative volumes of separable and nonseparable
states. Slater suggested that a natural measure to use in this context
was the normalized volume element of the Bures metric, and
implemented this idea in a number of investigations 
\cite{slaterpreprint,slater2,slaterJPA}. 
Numerical analyses conducted in \cite{slaterJPA} indicated that use of the
Bures metric led to {\it greater}
 probabilities of separability than did the
use of any other monotone metrics. (Also, at this stage 
of theoretical development,
the Bures metric appears to be more amenable computationally, due in 
particular to the work of Dittmann \cite{ditt2}.)

The  conjecture (\ref{e1}) 
concerning the {\it a priori} Bures/SD 
probability of
separability of two qubits, in turn,
rested upon one to the effect that the 
SD volume of separable (separable) qubit-qubit
pairs (representable by $4 \times 4$ density matrices) is
\begin{equation}\label{e2}
 V_{4}^{s} = 
 {\pi^6 \over 2 \cdot 3 \cdot 5 \cdot 7 \cdot 11} ={\pi^6 \over 2310}
\approx .416186
\end{equation}
(while the quasi-Monte Carlo estimate based upon 65 million points
in 15-dimensional space was .416302).

 This is expressible as the specific case $m=4$ of
the formula,
\begin{equation} \label{e3}
V_{m}^{s} = {\pi^k \over p_{k-1} \#} = 
{\pi^k \over \Pi_{i=1}^{k-1} p_{i}}, \qquad k={m (m-1) \over 2}, 
\end{equation}
 where  $p_{i}$ is the
$i$-th prime integer, and $p_{l} \#$ denotes the $l$-th ``primorial'' 
number \cite[seq. A002110 and refs. there]{sloane}. 
The primorial function satisfies the ``unexpected'' limit \cite{ruiz},
\begin{equation}
\mbox{lim} (p_{l} \#)^{1 \over p_{l}} = e, \quad l \rightarrow \infty.
\end{equation}

We report here 
for the qubit-qutrit case (representable
by $6 \times 6$ density matrices) that substitution of $m=6$ into this
formula, yields a result, 
\begin{equation} \label{e4}
V_{6}^{s} = {\pi^{15} \over p_{14} \#} = 
{\pi^{15} \over 2 \cdot 3 \cdot 5 \cdot 7 \cdot 11 \cdot 13 \cdot 17 \cdot 19 \cdot 23 \cdot 29 \cdot 31 \cdot37 \cdot 41 \cdot 43}  \approx 
2.19053 \cdot 10^{-9},
\end{equation}
 which is rather  closely approximated by our {\it two}  quasi-Monte
Carlo  estimates 
for the SD volume of the separable qubit-qutrit states.  
Combining results of Sommers and \,Zyczkowski \cite{sz} and of
Tilma and Sudarshan \cite{tilma}, in the manner discussed below,
 we are able to deduce that the SD volume of the
qubit-qutrit states ({\it both} separable and nonseparable/quantum in nature) is 
\begin{equation} \label{e5}
V_{6}^{s+n} = {\pi^{18} \over 2^{21}
 \cdot 3^4 \cdot 5^2 \cdot 7^2 \cdot 11 \cdot 13 \cdot 17}
\approx 1.75655 \cdot 10^{-6}.
\end{equation} 
Conjecturing the validity of the formula (\ref{e4}), then, easily 
implies a
further conjecture that
the {\it a priori} Bures/SD probability that a qubit and a
qutrit are separable is 
\begin{equation} \label{e6}
P_{6} = {V_{6}^{s} \over V_{6}^{s+n}} =  {2^{20} \cdot 3^3 \cdot 5 \cdot 7
\over 19 \cdot 23 \cdot 29 \cdot 31 \cdot 37 \cdot 41 \cdot 43 \pi^3}
\approx .00124706. 
\end{equation}
Earlier numerical analyses of ours \cite[Table 4]{slaterJPA}, for which the 
Euler-angle parameterization of $SU(6)$ \cite{tilma,tilma2} 
was not yet available, had been
marked by considerable instabilities. The best estimate there --- based on the
largest number of 
sampled density matrices --- gave a Bures/SD probability 
of qubit-qutrit separability
 of .00401427. We had also previously \cite[sec. 2.6.2]{slater2} 
reported the expression for the 
Bures volume element for  the {\it one}-dimensional ``hybrid'' 
Werner states composed of qubit-qutrit pairs.
It --- as well as the Bures 
volume element for the Werner states consisting of
qutrit-qutrit pairs \cite[sec. 2.5]{slater2} --- has a first-order 
pole at the pure state boundary, 
leading to a conclusion
that the corresponding Bures/SD probability of separability is {\it zero}.
The Bures/SD probability, on the other hand, that a Werner state composed
of a pair of qubits is separable is  
${1 \over 4}$ \cite[sec. 2.1.7]{slater2}.

It had already been quite firmly established  that 
\cite{slaterpreprint}
\begin{equation} \label{e7}
V_{4}^{s+n} = {\pi^8 \over 2^{4} \cdot 3 \cdot 5 \cdot 7} = 
{\pi^8 \over 1680} \approx 5.64794.
\end{equation}
The ratio $V{_4}^{s}/V_{4}^{s+n}$, then,
 gives the conjectured Bures/SD
probability (\ref{e1}), that is $8 /11 \pi^2$, that a pair of qubits is separable 
\cite{slaterpreprint}.
It is certainly intriguing
that the numbers 1680 and 2310, appearing in the denominators
of $V_{4}^{s+n}$ and $V_{4}^{s}$, have been conjectured by E. L\'abos
 \cite[seq. A064377, cf. A064375]{sloane}  to be 
the two {\it largest} integers ($l$), 
 having the property  
that the sum of
the {\it fourth} powers of the divisors of $l$ exceeds the {\it fifth}
 power of the 
number 
[Euler's totient function] of 
positive integers
relatively prime to $l$.  We sought to ascertain if there is any
generalization of this phenomenon to the qubit-qutrit case. We did find
that the sum of the $k$-th powers of the divisors of $14 \#$
does exceed the $(k+1)$-st  power of the 
corresponding Euler totient function for $k=19$, but not for $k<19$, and have
been unable to locate any larger numbers 
than $14 \#$ for which this holds.

The methodology we pursue here for the general
(15-dimensional) qubit-qutrit case 
is essentially the same as in 
the qubit-qubit analysis \cite{slaterpreprint}.
That is, we parameterize the $m \times m$ $(m=6)$ 
density matrices ($\rho$) using the newly-developed Euler
angle parameterization of $SU(m)$ \cite{tilma,tilma2}. We  
 linearly transform the 
$m^2-1 = 35$ variables
(30 Euler angles plus 5 spherical angles parameterizing the simplex of 
eigenvalues
of $\rho$), so that they range over a 35-dimensional hypercube, that is
all 35 transformed variables have values lying within the unit interval
[0,1].
Then, using advanced quasi-Monte Carlo procedures (scrambled Halton
sequences) \cite{okten}, we numerically integrate over this hypercube
the corresponding SD volume element, that is, the product
\cite[eq. (24)]{mjw},
\begin{equation}
\mbox{d} V^{s+n}_{6} = 
\mbox{d} D_{6} \mbox{d} H_{6}, 
\end{equation}
of the {\it conditional}
SD volume element $\mbox{d} D_{6}$ 
over the 5-dimensional {\it simplex} 
spanned by the six eigenvalues --- a density matrix, 
of course, having {\it unit} trace ---  and 
the ``truncated'' Haar volume element  $\mbox{d} H_{6}$ 
for the 30 Euler angles 
\cite{tilma}.

Simultaneously, in our calculations, 
we check if the particular $6 \times 6$ 
density matrix 
($\rho$)
parameterized 
by the  sampled point in the 35-dimensional space is separable
or not. We do this by determining whether or not 
 the partial transpose 
of $\rho$ has any negative eigenvalues. 
In fact, in contrast to our previous study of
the $4 \times 4$ case \cite{slaterpreprint}, we compute
 {\it two} forms of partial transpose \cite[eq. (21)]{Horodecki}: (1) 
individually transposing in place the {\it nine} $2 \times 2$
blocks of the $6 \times 6$ density matrix $\rho$ and; (2) individually
transposing in place 
the {\it four} $3 \times 3$ blocks. If one such
partial transpose has only nonnegative eigenvalues, 
then $\rho$ is separable
with respect to the corresponding partition \cite{Horodecki}. 
Let us designate the
partially transposed density matrices generated in these two
manners by $\rho^{T}_{9: 2 \times 2}$ and $\rho^{T}_{4: 3 \times 3}$.
(Of course, for the cases $m>6$ --- which we report upon briefly
below for $m=8,9$ --- a positive partial transpose is only a necessary
but not sufficient condition for separability \cite{Horodecki,andy}.)

We constructed a scrambled Halton sequence composed of 70 million points
lying within the 35-dimensional hypercube. (Our qubit-qubit analysis
\cite{slaterpreprint} was based on a 
sequence of 65 million points of a scrambled Halton sequence lying 
within a 15-dimensional hypercube.
Computations are, of course, considerably  more demanding in the qubit-qutrit
case, particularly in that 
we now calculate {\it two} forms of partial transpose for 
each point/density matrix. All the calculations in both studies
have been performed
in MATHEMATICA. The use of C or FORTRAN, at least for random number
generation, would, it appears, be somewhat beneficial timewise.) 
 
In Table I, we show the cumulative results of the estimation
process as the number of points of the scrambled Halton 
sequence is increased.
The table starts with
6 million points only because for fewer points we were not yet monitoring
the results at steps of one million, but only at rather irregular 
intervals.
\newpage
\begingroup
\squeezetable
\begin{table}
\begin{tabular}{r|r|r|r|r|r|r|r|r|r|r|}
(1) & (2) & (3) & (4) & (5) & (6) & (7) & (8) & (9) & (10) & (11) \\
\hline
$\#$  of points & $10^{6} D_{6} $ & $ H_{6}$ 
& (2)x(3) & $10^{6} V_{6}^{s+n} $ & $10^{9} V_{6}^{s} $ & 
$10^{9} V_{6}^{s}$ & $P_{6}$ & $P_{6}$ & 
avg. neg. & avg. log neg. \\
\hline
\hline
6,000,000 & 2.17311 & .819790 & 1.78149 & 1.63062  & 1.82876 & 2.81863 & .00121510 & .00172856 &  .111522 & .198113 \\
\hline
7,000,000 & 2.17163 & .818852 & 1.77824 & 1.59557  & 1.70201 & 2.63989 & .00106671 & .00165451 & 
.112394 & .199510  \\
\hline
8,000,000 & 2.16876 & .815219 & 1.76801 &  1.73148  & 1.73789 & 2.55169 & .00100370 & .00147371 & .113360 & .200963 \\
\hline
9,000,000 & 2.17067 & .813568 & 1.76599 & 
1.67732 & 1.60642  & 2.60212 & .00095773 & .00155136 & .113344 & .200921  \\
\hline
10,000,000 & 2.16961 & .811429 & 1.76048 & 1.65677 & 1.70955 & 2.53324 & .00103185 & .00152902 &
.114256 & .202339 \\
\hline
11,000,000 & 2.16896 & .809171 & 1.75506 & 
1.59914 & 1.73486 & 2.43431 & .00108487 & .00152226 & 
.114078 & .202011\\
\hline
12,000,000 & 2.16881 & .809422 & 1.75548 &1.66201 & 1.79158 & 2.27426 & .00107796 & .00136838 &.117223 &
.206971 \\
\hline
13,000,000 & 2.16720 & .808815 & 1.75287 & 1.67362 & 1.75997 & 2.16590 & .00105160 &.00129414 & .116156 & .205262 \\
\hline
14,000,000 & 2.16844 & .807586 & 1.75120 & 1.67940 & 1.67717 & 2.37734 &.00099867 & .00141559 & .115793 & .204657 \\
\hline
15,000,000 & 2.16762 & .808999 & 1,75360 & 1.65144 & 1.75328 & 2.36349 & .00106166 & .00143117 & .115576 & .204320 \\
\hline
16,000,000 & 2.16761 & .808868 & 1.75331 & 
1.65010 & 1.68137 & 2.29319 & .00101895 & .00138973 & .115228 & .203717 \\
\hline
17,000,000 & 2.16705 & .807704 & 1.75034 & 
1.64813 & 2.62922 & 2.22118 & .00159257 & 
.00134770 & .115345 & .203941 \\
\hline
18,000,000 & 2.16798 & .809567 & 1.75512 & 
1.72538 & 2.51953 & 2.12728 & .00146027 &
.00123293 & .114982 & .203534 \\
\hline
19,000,000 & 2.16822 & .808844 & 1.75375 & 
1.81214 & 2.51473 & 2.03554 & .00138771 & .00112328 & .113738 & .201610 \\
\hline
20,000,000 & 2.16833 & .809403 & 1.75505 & 1.79834 & 2.43912 & 2.09067 & .00135632  & .00116255  & .113124 & .200638 \\
\hline
21,000,000 & 2.16890 & .810723 & 1.75838 & 1.84181 & 2.38341 & 2.04151 & .00129406 & .00110842 & .111922 & .198704 \\
\hline
22,000,000 & 2.16864 & .810679  & 1.75807 & 1.83184 & 2.31055 & 2.05745 & .00126133 & .00112316 & .112238 & .199172 \\
\hline
 23,000,000 & {\bf 2.16845} & {\bf .809823} & {\bf 1.75606} & 
{\bf 1.84255} & {\it 2.33424} & 
{\it 2.07072} & {\it .00126685} & {\it .00112383} & {\it .112286} & 
{\it .199246} \\
\hline
24,000,000 & 2.16911 & .808835 & 1.75445 & 1.82218 & 2.34690 &
2.01018 & .00128797 & .00110318 & .112170 & .199042 \\
\hline
25,000,000 & 2.16897 & .809139 & 1.75500 & 1.82762 & 2.34520 & 2.04318 & .00128320 & .00111795 & .112399 & .199414 \\
\hline
26,000,000 & 2.16841 & .810069 & 1.75656 & 1.80906 & 2.34929 & 1.99359 & .00129862 & .00110201 & .112215 & .199109 \\
\hline
27,000,000 & 2.16854 & .809789 & 1.75606 & 1.79860 & 2.27939 & 1.93533 & .00126732 & .00107602 & .112600 & .199732 \\
\hline
28,000,000 & 2.16877 & .809939 & 1.75657 & 1.79198 & 2.90172 & 2.25790 & .00161929 & .00126000 & .112190 & .199061 \\
\hline
29,000,000 & 2.16921 & .808897 & 1.75467 & 1.77494 & 2.89076 & 2.19422 & .00162865 & .00123622 & .112218 & .199096 \\
\hline
30,000,000 & 2.16882 & .808270 & 1.75300 & 1.76758 & 2.82880 & 2.15317 & .00160038 & .00121815 & .111834 & .198465 \\
\hline
31,000,000 & 2.16945 & .808186 & 1.75332 & 1.75508 & 2.74992 & 2.10352 & .00156684 & .00119854 & .111656 & .198176 \\
\hline
32,000,000 & 2.16951 & .808231 & 1.75347 & 1.77407 & 2.71016 & 2.10345 & .00152766 & .00118566 & .112523 & .199533 \\
\hline
33,000,000 & 2.16916 & .807991 & 1.75266 & 1.77431 & 2.64319 & 2.06355 & .00114897 & .00116301 & .111969 & .198629 \\
\hline
34,000,000 & 2.16904 & .808520 & 1.75371 & 1.76566 & 2.58045 & 2.08714 & .00146147 & .00118208 & .111714 & .198225 \\
\hline
35,000,000 & 2.16903 & .808302 & 1.75323 & 1.77001 & 2.54634 & 2.04145 & .00143860 & .00115335 & .111903 & .198521 \\
\hline
36.000,000 & 2.16909 & .807965 & 1.75255 & 1.77289 & 2.49867 & 2.00118 & .00140937 & .00112877 & .111890 & .198519 \\
\hline
37,000,000 & 2.16953 & .808300 & 1.75363 & 1.77258 & 2.47890 & 1.97301 & .00139847 & .00111307 & .111893 & .198517 \\
\hline
38,000,000 & 2.16907 & .808002 & 1.75262 & 1.77682 & 2.50868 & 1.95478 &
.00141189 & .00110016 & .111587 & .198022 \\
\hline
39,000,000 & 2.16941 & .808093 & 1.75308 & 1.77754 & 2.45603 & 1.92508 & .00138170 & .00108300 & .111890 & .198501 \\
\hline
40,000,000 & 2.16947 & .808559 & 1.75414 & 1.77983 & 2.40611 & 1.89478 & .00135187 & .00106458 & .111828 & .198397 \\
\hline
41,000,000 & 2.16993 & .809146 & 1.75579 & 1.77556 & 2.36968 & 1.86274 & .00133461 & .00104910 & .111880 & .198448 \\
\hline
42,000,000 & 2.16914 & .808738 & 1.75427 & 1.77559 & 2.34147 & 1.95985 & .00131870 & .00110378 & .112031 & .198657 \\
\hline
43,000,000 & 2.16899 & .808551 & 1.75374 & 1.78547 & 2.29604 & 1.94252 & 
.00128596 & .00108796 & .111850 & .198369 \\
\hline
44,000,000 & 2.16861 & .809352 & 1.75517 & 1.78240 & 2.25845 & 1.91128 & .00126708 & .00107231 & .111720 & .198155 \\
\hline
45,000,000 & 2.16833 & .809135 & 1.75447 & 1.77038 & 2.26746 & 1.87635 & .00128078 & .00105986 & .111774 & .198237 \\
\hline
46,000,000 & 2.16820 & .809338 & 1.75481 & 1.76196 & 2.24469 & 1.84931 & .00127397 & .00104957 & .111685 & .198092 \\
\hline
47,000,000 & 2.16854 & .809032 & 1.75442 & 1.74799 & 2.18383 & 1.85180 & 
.00124934 & .00105939 & .111543 & .197868 \\
\hline
48,000,000 & 2.16852 & .809139  & 1.75463 & 1.75059 & 2.16945 & 1.85939 & .00123927 & .00106215 & .111492 & .197792  \\
\hline
49,000,000 & 2.16865 & .809302 & 1.75509 & 1.74786 & 2.16569 & 1.84002 & .00123905 & .00105273 & .111400 & .197654 \\
\hline
50,000,000 & 2.16870 & .809839 & 1.75630 & 1.74852 & 2.15184 & 1.82899 &
.00123067 & .00104602 & .111310 & .197523 \\
\hline
51,000,000 & 2.16902 & .810318 & 1.75760 & 1.75002 & 2.12687 & 1.80091 & 
.00121534 & .00102908 & .111426 & .197725 \\
\hline
52,000,000 & 2.16896 & .810052 & 1.75697 & 1.74811 & 2.15106 & 1.79672 & .00123051 & .00102780 & .111131 & .197226 \\
\hline
53,000,000 & 2.16804 & .809509 & 1.75505 & 1.75131 & 2.13952 & 1.77231 & .00122167 & .00101199 & .110985 & .196998 \\
\hline
54,000,000 & 2.16801 & .809300 & 1.75457 & 1.76323 & 2.14336 & 1.74553 & 
.00121559 & .00098965 & .111036 & .197100 \\
\hline
55,000,000 & 2.16792 & .808631 & 1.75304 & 1.77523 & 2.16057 & 1.72051 & .00121707 & .00096917 & .110937 & .196945 \\
\hline
56,000,000 & 2.16804 & .808953 & 1.75384 & 1.77400 & 2.16228 & 1.69447 & .00121887 & .00095516 & .110961 & .196970 \\
\hline
57,000,000 &  2.16839 & .808805 & 1.75380 & 1.76956 & 2.15098 & 1.73970 &
.00121554 & .00098312 & .110807 & .196720 \\
\hline
58,000,000 & 2.16870 & .808821 & 1.75409 & 1.76301 & 2.13958 & 1.73100 & .00121360 & .00098184 & .110919 & .196902 \\
\hline
59,000,000 & 2.16841 & .808892 & 1.75401 & 1.76677 & 2.10854 & 1.72040 & .00119344 & .00097375 & .110741 & .196598 \\
\hline
60,000,000 & 2.16813 & .808575 & 1.75310 & 1.76185 & 2.07877 & 1.88695 & .00117988 & .00107101 & .110813 & .196721 \\
\hline
61,000,000 & 2.16801 & .808273 & 1.75234 & 1.75346 & 2.08729 & 1.90074 & 
.00119038 &  .00108399 & .110825 & .196741 \\
\hline
62,000,000 & 2.16792 & .808387 & 1.75252 & 1.75237 & 2.08725 & 1.89066 & .00119110 & .00107892 & .110735   &.196597  \\
\hline
63,000,000 & 2.16812 & .808415 & 1.75274 & 1.75189 & 2.08029 & 1.88303 & 
.00118745 & .00107486 & .110899 & .196864 \\
\hline 
64,000,000 & 2.16815 & .807768 & 1.75136 & 1.74927 & 2.05540 & 1.90847 & 
.00117500  & .00109101 & .110868 & .196826 \\
\hline
65,000,000 & 2.16821 & .807592 & 1.75103 & 1.74516 & 2.06743 & 1.89593 & .00118467 & .00108639 & .110955 & .196961 \\
\hline
66,000,000 & 2.16855 & .807946 & 1.75207 & 1.74917 & 2.04261 & 1.91503 & 
.00116776 & .00109482 & .110935 & .196925 \\
\hline
67,000,000 & 2.16862 & .808183 & 1.75264 & 1.75094 & 2.02017 & 1.89124 & 
.00115376 & .00108012 & .110844 & .196781 \\
\hline
68,000,000 & 2.16855 & .807698 & 1.75153 & 1.75375 & 2.01899 & 
1.87126 & .00115124 & .00106701 & .110742 & .196615 \\
\hline
69,000,000 & 2.16851 & .807844 & 1.75182 & 1.74229 & 2.00726 & 1.85930 
&   .00115208 & .00106716 & .110747 & .196624 \\
\hline
70,000,000 & 2.16855 & .807921 & 1.75202 & 1.74533 & 2.04247 
& 1.84188 & .00117025 & .00105532 & .110584 & .196355 \\
\hline
\hline
{\it conj.}{\bf /act.} & {\bf 2.16914} & {\bf .809794} & {\bf 1.75655} & 
{\bf 1.75655} & {\it 2.19053} & {\it 2.19053} &
{\it .00124706} & {\it .00124706} & --- & ---\\
\end{tabular}
\caption{Cumulative quasi-Monte Carlo estimates, based on scrambled Halton 
sequences, of $D_{6}, H_{6},V_{6}^{s+n}, 
V_{6}^{s},
P_{6}$
and the two 
entanglement measures, average negativity and average log negativity, 
for qubit-qutrit pairs. The best approximation to the truncated Haar 
volume $H_{6}$ occurs
for 23,000,000 points, at which the average of the two estimates of
$10^9 V_{6}^{s}$ is 2.20248, close to the conjectured value of
2.19053.}
\end{table}
\endgroup
\nopagebreak
 The left-positioned estimates in the table
of $V_{6}^{s}$ and $P_{6}$ are  based on the partial 
transpose $\rho^{T}_{4:3 \times 3}$
and the right-positioned estimates  
on $\rho^{T}_{9:2 \times 2}$. (Of the 68  million $6 \times 6$ 
density matrices 
generated on the basis of the scrambled Halton sequence, 3,393,933
of them were separable with respect to this first form
of partial transposition, and 2,928,456  with respect to the second.) 
In the fourth column --- which can be 
compared with the fifth --- is the simple product
of the entries in the second and third columns. The last line 
of the table contains
our (scaled) known or conjectured values. The best tabulated
approximation,
 .809823,  to the truncated Haar volume 
$H_{6} \approx .809794$
occurs for 23,000,000 points of the scrambled Halton sequence.  
In retrospect, 
it would have been interesting to monitor
our computations at those stages, not just {\it regular}
intervals of one million,  at which $V^{s+n}_{6}$, was particularly
well-approximated, since presumably at these stages we would also tend to
obtain relatively accurate estimates of $V^{s}_{6}$, among other 
variables of interest. 
Analogous
monitoring 
approaches would also be of potential interest in the qubit-qubit case
\cite{slaterpreprint} (where both $H_{4}$ and $D_{4}$, and thus
$V^{s+n}_{4}$, have been known), 
as well as in the qubit-qubit-qubit and qutrit-qutrit
cases 
briefly analyzed at the conclusion of this paper (where, 
now due to the quite recent results of Sommers and \.Zyczkowski
\cite{sz}, $D_{8}$ and $D_{9}$ are known, in addition to $H_{8}$ and
$H_{9}$).

To determine  the 
value of the SD volume ($V^{s+n}_{6}$), we first ascertained
 (through simple exact 
integration) the value of 
the truncated Haar volume ($H_{6}$), 
that is,
\begin{equation} \label{e8}
H_{6}={\pi^{15} \over 2^{18} \cdot 3^{3} \cdot 5} 
\approx .809794.
\end{equation}
We also observe that
\begin{equation}
H_{4} = { \pi^6 \over 2^5 \cdot 3}  \approx 10.0145.
\end{equation}
Both these  measures are  ``truncated'' forms of the Haar measure for
$SU(m)/Z_{m}$ because $(m-1)$ of the $m(m-1)$ Euler
angles, associated with {\it diagonal} Lie
generators
become irrelevant [that is, ``drop out''] 
in the formation of $\rho$.
Since the points of the (``low discrepancy'') 
scrambled Halton sequence are devised so as to
closely approximate the {\it uniform} distribution over the hypercube, it is
natural to expect that the distribution of points over {\it sub}hypercubes, 
as well, 
should also be close to uniform. (It might also be of interest to conduct
a parallel analysis, not using a 35-dimensional hypercube, 
but rather a 5-dimensional 
one  with an independently generated  30-dimensional hypercube.) 
Sommers and \.Zyczkowski have shown \cite[eq. (4.11)]{sz}
 (rather late in the course
of this research) that, in general (we adopt 
their [Bures] formula to the SD case),
\begin{equation}
D_{n} = 
{\pi^{n \over 2} \Pi^{n+1}_{i=1} \Gamma(i) \over
 \Gamma({n^2 \over 2})},
\end{equation}
so
\begin{equation} \label{e9}
D_{6} = {\pi^3 \over 2^{3} \cdot 3 \cdot 5 \cdot 7^2 \cdot 11
\cdot 13 \cdot 17} \approx 2.16914 \cdot 10^{-6}.
\end{equation}
(This value had been {\it conjectured} in \cite{slaterhall}.)
Also  \cite{slaterpreprint},
\begin{equation}
D_{4} ={2 \pi^2 \over 5 \cdot 7} \approx .563977.
\end{equation}

The product of $H_{6}$  and $D_{6}$ gives us (\ref{e5}), or more
generally, we have
$H_{m} D_{m} = V_{m}^{s+n}$.
In general, we have for the SD volume element restricted to the
diagonal density matrices \cite[eq. (25]{mjw},
\begin{equation} \mbox{d} D_{m} = { \mbox{d} \lambda_{1} \ldots \mbox{d} 
\lambda_{m-1} \over \sqrt{\Pi_{i=1}^{m} \lambda_{i}}} \Pi_{1 \leq i < j}^{m}
{4 (\lambda_{i} -\lambda_{j})^{2} \over (\lambda_{i} +\lambda_{j})},
\end{equation}
where the $\lambda$'s are the eigenvalues of $\rho$.

Our numerical estimates that a $6 \times 6$ density matrix is separable
with respect to {\it both} forms of partial transposition employed here
were much lower than that either, independent of the other, is separable,
 suggesting a rough independence
of these properties.

In Fig.~1, on the basis of the results presented in Table I, 
 we show the deviations from the conjectured value of 
$10^9 V_{6}^{s}$ of the average of the two estimates (based on the two
different forms of partial transposition).
\newpage
\begin{figure}
\centerline{\psfig{figure=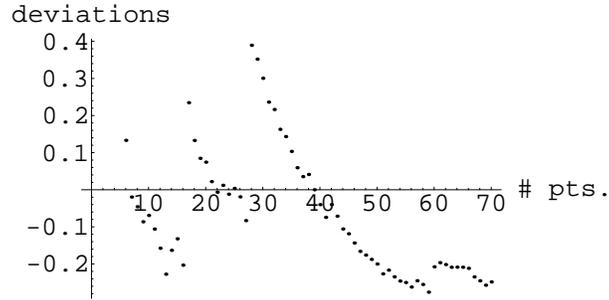}}
\caption{Deviations of the average of the two estimates --- based on the
two forms of partial transposition --- of 
$10^9 V^{s}_{6}$ 
from its conjectured value of 2.19053. Numbers of points are in millions.
Data correspond to the averages of columns (6) and (7) of Table I. The 
best approximation to the truncated Haar volume 
$H_{6}$ occurs for 23,000,000 points, at which the
average of the two estimates of $10^9 V_{6}^{s}$ is 2.20248, close to
the conjectured value of 2.19053.}
\end{figure}

In Fig.~2 are displayed the deviations from the actual/known value
of $H_{6}$ obtained during the quasi-Monte Carlo procedure.
\begin{figure}
\centerline{\psfig{figure=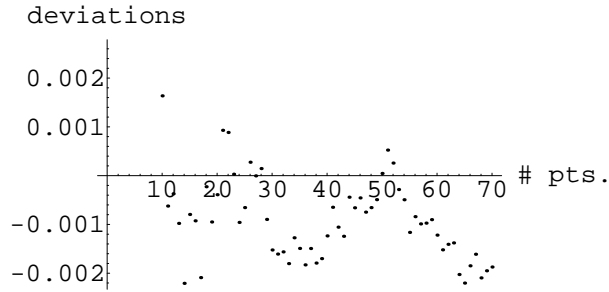}}
\caption{Deviations from the actual/known value of $H_{6} = {\pi^{15} \over 
35389440} \approx .809794$ obtained during the quasi-Monte Carlo procedure. Number of points
are in millions. Data correspond to column (3) of Table I. The best
approximation occurs at 23,000,000 points}
\end{figure}

In Fig.~3 are shown the deviations multiplied by $10^{6}$ 
from the known  value of $D_{6} \approx 2.16914  \cdot 10^{-6}$, that is (\ref{e9}) obtained during the quasi-Monte Carlo procedure.
\begin{figure}
\centerline{\psfig{figure=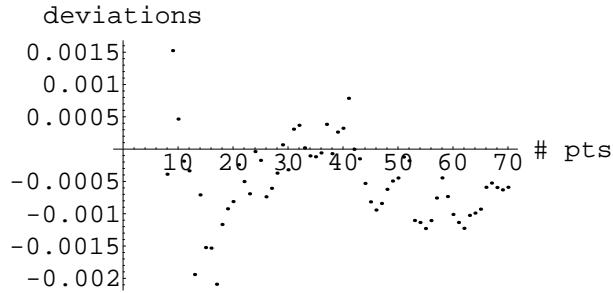}}
\caption{Deviations from the actual/known value of $10^6 D_{6} 
\approx 2.16914 $ obtained during the
quasi-Monte Carlo procedure. Number of points are in millions. Data
correspond to column (2) of Table I}
\end{figure}
\newpage

We also estimated 
the SD area ($A_{6}^{s}$) 
of the {\it 34}-dimensional boundary between the separable and 
nonseparable qubit-qutrit states to be
$\approx 1.094257 \cdot 10^{-6}$. This was accomplished by generating 
751,000 points of a 34-dimensional random Halton sequence. 
For each of 629,819 of these points, we were able to find  value(s) of the 
35-$th$ parameter for 
which the partial transpose of the so-parameterized density matrix would 
have a zero 
eigenvalue. 
In this way, 1,308,734 density matrices on the separable/nonseparable 
boundary were obtained.  
In Table II, we display these results together with a few other intermediate
ones.
\begin{table}
\begin{tabular}{r|r|r|r|}
\# of sampled points & \# of feasible sampled points & \# of density 
matrices on sep./nosep. boundary
& $10^6 A^{s}_{6}$ \\
\hline
35,000 & 29,323 & 60,757 & 1.294234 \\
\hline
258,000 & 216,210 & 449,113 & 1.156524 \\
\hline
550,000 & 465,156 & 966,833 & 1.062530 \\
\hline
751,000 & 629,819 & 1,308,734 & 1.094257 \\
\end{tabular}
\caption{Cumulative quasi-Monte Carlo estimates, based on scrambled Halton 
sequences, of the SD area $A^{s}_{6}$ 
of the 34-dimensional
boundary of the separable qubit-qutrit states}
\end{table}

We had also in \cite{slaterpreprint}
obtained, on the basis of some 11,800,000 points,
 an estimate $A^{s}_{4} \approx 1.75414$ of the analogous
14-dimensional boundary in  the qubit-qubit case.
Also, for the SD boundary area of the convex set of qubit-qutrit pairs,
we found $A_{6}^{s+n} \approx .00001874312$, while 
the comparable quantity in the qubit-qubit case was \cite{slaterpreprint},
\begin{equation}
 A_{4}^{s+n} = {2 \cdot 71 \pi^7 \over 3^3 \cdot 5 \cdot 7 \cdot 13} 
 \approx 34. 911000.
\end{equation}

The scalar curvature of the 35-dimensional Riemannian 
manifold of qubit-qutrit
states endowed with the Bures metric assumes its minimum, 3080, at the 
fully mixed state \cite{Jochen}. 
If, in addition, to the scalar curvature, the Ricci curvature were also
bounded below, in particular, by $m^2-1=34$ (which we 
unfortunately at this time do {\it not} specifically know is or is
not the case),
then we could 
directly apply the
``Levy-Gromov'' isoperimetric inequality \cite[p. 520]{gromov}.
In this case, the ratio (.616806) of the area $A_{6}^{s} \approx 1.09426 
\cdot 10^{-6}$ 
of the 
boundary
of separable states to the volume $V^{s+n} \approx 1.77407 10^{-6}$
 would be {\it greater} than
the ratio ($w$) of $s(\alpha)$ to the volume $\tilde{V} \approx 
3.34529 \cdot 10^{-7}$ of
a unit ball in 35-dimensional space. Now $\alpha$ itself is the
ratio $V^{s} /V^{s+n} \approx .0013566$ and $s(\alpha)$ is then 
the area
of the boundary of a ball in 35-dimensional space having a volume
equal to $\alpha \tilde{V}$. This gives us $w \approx .05734$, 
so the desired
Levy-Gromov 
inequality holds, since $.616806 > .05734$, 
which is consistent with, but not probative 
that the Ricci curvature
for the qubit-qubit states endowed with the Bures metric nowhere
assumes a value less than 34. (In the qubit-qubit case 
\cite{slaterpreprint}, the analogous
evidence indicated that the Levy-Gromov isoperimetric inequality did
{\it not} hold in that setting, so the Ricci curvature had to assume
a value somewhere less than 14.)
The lower bound on the Ricci curvature 
required for the potential application (saturation) 
of an isoperimetric inequality
would --- based on our estimates of $V^{s+n}_{6}, V^{s}_{6}$ and
$A^{s}_{6}$ --- be raised from 34 to $34 \kappa = 3934.06$ by using a 
certain extension 
\cite[Thm. 6.6]{chavel} \cite{gromov2} of
the Levy-Gromov result. (Here $\kappa$ fulfills the role of 
the constant
sectional curvature of a 35-dimensional sphere of radius 
$1 / \sqrt{\kappa}$.)

We also  computed the average entanglement --- as measured by the
{\it negativity} (the absolute value of the sum of the negative
eigenvalues of the partial transpose) and the {\it logarithmic negativity}
(the log of the sum of twice the negativity plus 1) 
\cite{vidalwerner} --- with respect to the 
Bures/SD measure for the two forms of partial transposes.
The average values across these two forms are reported in Table I.
In the qubit-qubit analysis \cite{slaterpreprint}, the average
(Bures/SD) negativity found was .177162 and the average {\it concurrence},
.197284.

We would like to argue that the potential applicability of the
formula (\ref{e3}) here to the qubit-qutrit case, adds some additional
weight to the plausibility of its applicability to the qubit-qubit case,
as presented in \cite{slaterpreprint}.
Of course, one might argue, in general, 
that the numerical evidence we present
for our conjectures is far from fully convincing. We 
certainly do not disagree, but
would like to emphasize that it is very difficult (``the curse of
dimensionality'') to achieve 
substantial accuracy
in high-dimensional spaces. Additionally, we are trying to evaluate
an apparently relatively {\it small} effect, that is, the SD/Bures 
probability that a qubit-qutrit pair is separable.

Ongoing related projects, with similar objectives, 
involve analyses of arbitrary states of three
qubits, and of two qutrits (based on Euler angle parameterizations, 
respectively,
of $SU(8)$ and $SU(9)$ \cite{tilma}), 
as well as of the more specialized Eggeling-Werner tripartite
states \cite{egg,slateregg}. 

In fact, for the two-qutrit case, we have for the truncated Haar volume,
\begin{equation}
H_{9} = {\pi^{36} \over 2^{51} \cdot 3^9 \cdot 5^4 \cdot 7^2} =
\approx 5.81699 \cdot 10^{-7}.
\end{equation}
On the basis of 1,972,000 points so-far computed 
of a scrambled Halton sequence over
the corresponding 
80-dimensional hypercube, we obtained an estimate of $H_{9}$ of
$6.01879   \cdot 10^{-8}$. 
(A standard ``rule-of-thumb'' asserts that $2^{80} \approx 1.20893 
\cdot 10^{34}$ 
points would be appropriate.) The mean negativity was .0936498 and the
mean log negativity was .169715. The estimate of $V_{9}^{s+n}$ was
$6.84438   \cdot 10^{-34}$ and of $V_{9}^{p}$ (the volume of states with 
positive partial transposes, which in general need not be separable 
for $m>6$ \cite{Horodecki}) was $8.957256 \cdot 10^{-52}$
(while a possible application of 
formula (\ref{e3}) would give $2.50963 \cdot 10^{-30}$).
The ratio of these two 
figures gives us an estimate of the Bures/SD probability
of having a positive partial transpose of $1.3087  \cdot 10^{-18}$.

In the {\it three}-qubit case, we have for the truncated Haar volume,
\begin{equation}
H_{8} = {\pi^{28} \over 2^{37} \cdot 3^{7} \cdot 5^{3} \cdot 7} 
 \approx .000316395.
\end{equation}
On the basis of 2,450,000 points (of a scrambled Halton sequence in
63-dimensional space), 
we obtained an estimate of $H_{8}$
of .000442676.
We estimated the mean negativity to be .098168 and
the mean log negativity to be .17503. (These and subsequent mean values, 
when appropriate, 
were obtained by further averaging of three sets of results --- one for each 
possible type of 
partial transposition.) The SD volume of states with
positive partial transposes
was estimated to be $1.94423  \cdot 10^{-30}$, with formula (\ref{e3})
giving $3.46944 \cdot 10^{-27}$, while the SD volume of all
states was estimated to be $1.041545  \cdot 10^{-21}$.
Using the three ``witnesses'' described in \cite[eqs. (6)-(8)]{acin},
we obtain estimates of lower bounds on
 the Bures/SD probability of being a GHZ state of $3.26363 \cdot 10^{-39}$,
and of being a W-state of both $8.68173 \cdot 10^{-21}$ and 
$2.92036  \cdot 10^{-12}$.

\acknowledgments

I would like to express appreciation to the 
Kavli Institute for Theoretical
Physics for computational  support in this research, to T. Tilma
for supplying the author with his analysis \cite{tilma2}, 
and to K. \.Zyczkowski for certain helpful comments.

\end{document}